\documentclass[journal,draftcls,onecolumn,letterpaper,twoside,11pt]{IEEEtran}

\usepackage[cmex10]{amsmath}
\usepackage{latexsym,amssymb,cite,bbm,epsf,amsthm,subfigure,algorithm,algorithmic}
\usepackage{graphicx}
\usepackage{enumerate}
\usepackage{mathrsfs}
\usepackage{epstopdf}
\usepackage{bbm}

\newcommand{\bc}{\begin{center}}
\newcommand{\ec}{\end{center}}
\newcommand{\be}{\begin{equation}}
\newcommand{\ee}{\end{equation}}
\newcommand{\ba}{\begin{array}}
\newcommand{\ea}{\end{array}}
\newcommand{\bea}{\begin{eqnarray}}
\newcommand{\eea}{\end{eqnarray}}
\newcommand{\bal}{\begin{align}}
\newcommand{\eal}{\end{align}}
\newcommand{\ei}{\end{itemize}}
\newcommand{\bi}{\begin{itemize}}
\newcommand{\bfi}{\begin{figure}}
\newcommand{\efi}{\end{figure}}
\newcommand{\MB}{\left[\begin{array}}
\newcommand{\ME}{\end{array}\right]}
\newcommand{\nn}{\nonumber}

\newtheorem{thm}{Theorem}

\newtheorem{cor}{Corollary}

\newcommand{\Exp}{\mathsf{E}}
\newcommand{\bExp}{\bar{\mathsf{E}}}
\newcommand{\Pro}{\mathsf{P}}
\newcommand{\bPro}{\bar{\mathsf{P}}}
\newcommand{\Hyp}{\mathsf{H}}

\newcommand{\cF}{\mathcal{F}}
\newcommand{\cG}{\mathcal{G}}

\newcommand{\cN}{\mathcal{N}}

\newcommand{\cC}{\mathcal{C}}

\newcommand{\bN}{\mathbb{N}}

\newcommand{\ind}[1]{\mathbbm{1}_{\{#1\}}}   

\newcommand{\ignore}[1]{{}}

\usepackage{tikz}

\begin{document}

\title{Sequential Joint Spectrum Sensing and Channel Estimation for Dynamic Spectrum Access}

\author{Yasin~Y{\i}lmaz\IEEEauthorrefmark{1}\footnote{\IEEEauthorrefmark{1}Electrical Engineering Department, Columbia University, New York, NY 10027.},\;\;
        Ziyu~Guo\IEEEauthorrefmark{2}\footnote{\IEEEauthorrefmark{2}National Mobile Communications Research Lab., Southeast University, Nanjing 210096, China.},\;\;
        and \, Xiaodong Wang\IEEEauthorrefmark{1}}

\maketitle

\begin{abstract}
Dynamic spectrum access under channel uncertainties is considered. With the goal of maximizing the secondary user (SU) throughput subject to constraints on the primary user (PU) outage probability we formulate a joint problem of spectrum sensing and channel state estimation. The problem is cast into a sequential framework since sensing time minimization is crucial for throughput maximization. In the optimum solution, the sensing decision rule is coupled with the channel estimator, making the separate treatment of the sensing and channel estimation strictly suboptimal. Using such a joint structure for spectrum sensing and channel estimation we propose a distributed (cooperative) dynamic spectrum access scheme under statistical channel state information (CSI). In the proposed scheme, the SUs report their sufficient statistics to a fusion center (FC) via level-triggered sampling, a nonuniform sampling technique that is known to be bandwidth-and-energy efficient. Then, the FC makes a sequential spectrum sensing decision using local statistics and channel estimates, and selects the SU with the best transmission opportunity. The selected SU, using the sensing decision and its channel estimates, computes the transmit power and starts data transmission. Simulation results demonstrate that the proposed scheme significantly outperforms its conventional counterparts, under the same PU outage constraints, in terms of the achievable SU throughput.
\end{abstract}


\section{Introduction}

Addressing the well-known problem of spectrum utilization scarcity in current wireless networks, the cognitive radio (CR) technology employs a hierarchical spectrum access model consisting of primary users (PUs) and secondary users (SUs) \cite{Zhao07}. In this model, both PUs and SUs are able to access a same band with a higher priority for PUs.
The spectrum sharing between PUs and SUs can be realized in an \emph{underlay} fashion, which allows SUs to coexist with PUs without sensing the spectrum band. Thus, SUs are blind to the idle state of PUs (spectrum holes), resulting in a worst-case assumption that PUs use the band all the time. As a result, SUs can coexist only with severe constraints on the transmission power in order to protect the quality of service (QoS) of PUs. Focusing on the analysis of underlay spectrum access, \cite{Kang09_1,Musavian09,Kang11} derive fading channel capacities and optimum power allocation strategies for SUs.
In contrast to underlay, the \emph{opportunistic access} approach permits the existence of SUs only when PUs are idle, i.e., no coexistence. Hence, in this approach there is no harsh constraints on the SU transmission power. Instead, an effective spectrum sensing scheme is needed \cite{Liang08,Chen08,Fan11}. In \cite{Liang08,Chen08} the SU throughput is maximized while satisfying the PU QoS constraints.

Methods for combining the underlay and opportunistic access approaches have also been proposed, e.g., \cite{Devroye06,Kang09_2,Zhang09,Chen13}. In such combined methods, the SU senses the spectrum band, as in opportunistic access, and controls its transmit power using the sensing result, which allows SU to coexist with PU, as in underlay. While deriving the power control function, the average or peak constraints on SU transmit power and PU interference level are imposed \cite{Zhang09,Chen13}. In this paper, we propose such a combined method under the peak interference and power constraints.
In spectrum access methods it is customary to assume perfect channel state information (CSI) at the SU, e.g., \cite{Kang09_1,Musavian09,Kang11,Devroye06,Kang09_2,Zhang09,Chen13}. That is, the perfect CSI of SU channels (and even PU channels) can be made available to the SU. The quantized CSI case is treated in \cite{Chen13}. However, how to obtain  the  CSI in the process of dynamic spectrum access has not been addressed. We consider the problem of joint spectrum sensing and channel estimation in this work.

For such a joint problem, a straightforward solution is to treat the two subproblems separately by using the optimum solution for each subproblem. More specifically, one can use the likelihood ratio test (LRT) for spectrum sensing and the minimum mean square error (MMSE) estimator for channel estimation to solve the joint problem. However, as shown in \cite{Moustakides12,Yilmaz13_jde}, treating each subproblem separately and solving it optimally does not necessarily result in the optimum overall performance. In \cite{Moustakides12,Middleton68,Fredriksen72}, optimum solutions to different formulations of the joint detection and estimation problem are given under the fixed-sample-size framework. More recently, in \cite{Yilmaz13_jde} a \emph{sequential joint detection and estimation} problem is considered, and the optimum solution is given, where the decision rule is a function of the estimator, making the separate treatment strictly suboptimal. The sequential framework ideally suits the goal of maximizing the SU throughput in dynamic spectrum access. In particular, it is desirable to perform reliable sensing as soon as possible to let the SU transmit data as long as possible, leading to higher throughput. Indeed, in the sequential framework the sensing time is minimized.
Here we propose a dynamic spectrum access method based on sequential joint spectrum sensing and channel estimation.

Pilot signals are often used in channel estimation, e.g., \cite{Yue04,Li00}, and also in spectrum sensing, e.g., \cite{Tang05,Sahai06,Mishra07}. We similarly propose to make use of the pilot signals transmitted for PU communications to jointly sense and estimate the channels linked to  the SU. In a cognitive radio network, multiple SUs can cooperate to sense the spectrum by sharing their local information either over a fusion center (FC) or directly with other SUs. For such a decentralized system bandwidth and energy-efficient scheme is required for information transmission and processing. Recently, in a series of papers \cite{Fellouris11,Yilmaz12,Yilmaz13_est}, it is shown that a nonuniform sampling technique called \emph{level-triggered sampling} is an ideal fit for distributed information transmission and processing. This is because it enables highly accurate recovery at the FC by transmitting only a single bit per sample. Furthermore, it allows for complete asynchrony among SUs, a highly desirable feature in distributed systems, and censors uninformative local information. Due to its attractive features we use level-triggered sampling in the proposed dynamic spectrum access scheme to enable cooperation between SUs.

The remainder of the paper is organized as follows. In Section \ref{sec:system}, we formulate the problem and briefly discuss the conventional spectrum access methods. Then, in Section \ref{sec:sjde} the sequential joint spectrum sensing and channel estimation problem is introduced and the optimum solution is given. The proposed cooperative spectrum access scheme is given in Section \ref{sec:dsa}, and simulation results comparing its performance with other schemes are provided in Section \ref{sec:sim}. Finally, the paper is concluded in Section \ref{sec:conc}.

\section{System Descriptions}
\label{sec:system}

\begin{figure}[t]
\centering
\includegraphics[scale=0.5]{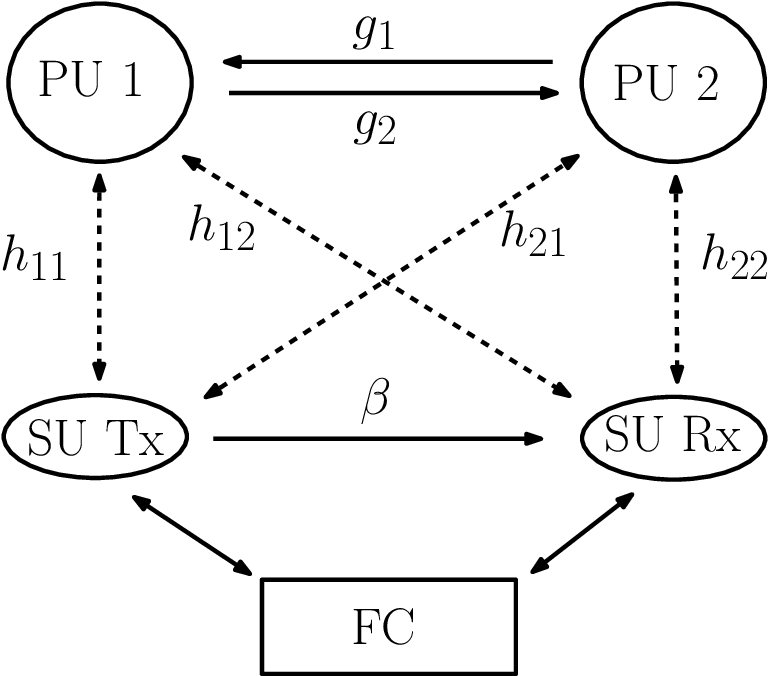}
\caption{The cognitive radio system under consideration.}
\label{fig:system}
\end{figure}

Consider a cognitive radio network consisting of a primary user (PU) pair, a secondary user transmitter (SU Tx) and receiver (SU Rx), and a fusion center (FC), as shown in Fig. \ref{fig:system}, where the PU pair can simultaneously communicate to each other through full duplexing. Although no direct communication takes place between the PUs and the SUs, interference to the PU communications occurs through the cross links, represented by dashed lines in Fig. \ref{fig:system}. The FC facilitates cooperation among the SUs, and it can be either a dedicated entity or one of the SUs. The channel, i.e., cross link, between PU $i$ and SU Tx is represented by a complex random coefficient, i.e., channel gain, $h_{i1}$. Similarly the complex random coefficient $h_{i2}$ denotes the channel gain between the PU $i$ and SU Rx. We assume Rician fading channels, i.e., the real and imaginary parts of $h_{ik}$, $\Re(h_{ik})$ and $\Im(h_{ik})$, are independent and identically distributed (i.i.d.) as $\cN(\mu_{ik},\sigma_{ik}^2),\forall i,k$, with $\mu_{ik}=0$ corresponding to Rayleigh fading channels. Moreover, $\{h_{ik}\}$ are assumed to be independent, but they are in general not identically distributed with different means and variances.

\subsection{Problem Formulation}

As a fundamental requirement in cognitive radio systems, the SUs should not cause degradation in the quality of service (QoS) to the PUs. In other words, the interference from the SUs to the PUs must be kept below some maximum tolerable levels. Under such interference constraints, a natural objective is to maximize the SU throughput, i.e., the average bit-rate
 of SU Tx. Hence, assuming Gaussian noise in channels between the PUs and also between the SUs we aim to solve the following optimization problem
\be
\label{eq:opt_pro}
	\max_{P(h_{11},h_{21}) \leq P_{\text{max}}} \left\{ \ba{ll} \log\left(1+\frac{|\beta|^2 P(h_{11},h_{21})}{N_0}\right) & \text{if} ~~\Hyp_0 \\
	\ba{l} \log\left(1+\frac{|\beta|^2 P(h_{11},h_{21})}{N_0 + |h_{12}|^2 Q_1 + |h_{22}|^2 Q_2}\right) \\ \text{s.t.} ~~ |h_{11}|^2 P(h_{11},h_{21}) \leq I_1 ~~ \text{and}~~ |h_{21}|^2 P(h_{11},h_{21}) \leq I_2 \ea & \text{if} ~~\Hyp_1 \ea \right.,
\ee
where $P(h_{11},h_{21})$ is the transmit power of SU Tx, constrained by the maximum power $P_{\text{max}}$, and is a function of the channel gains $h_{11},h_{21},$ between SU Tx and the PUs; $\beta$ and $N_0$ are the channel gain and the variance of the Gaussian noise, respectively, between SU Tx and SU Rx; $Q_1$ and $Q_2$ are the transmit powers for PUs; and $I_1$ and $I_2$ are the maximum tolerable interference powers at PUs, which are determined by the PU outage constraints. The null hypothesis $\Hyp_0$ and the alternative hypothesis $\Hyp_1$ correspond to the absence and presence of PU communication, respectively. More specifically, $Q_1=Q_2=0$ under $\Hyp_0$, whereas $\max\{Q_1,Q_2\}\not=0$ under $\Hyp_1$.
\ignore{
For simplicity, we assume only one SU selected by the FC transmits on the band licensed to the PU pair. This single band framework can be easily extended to the multiple band case where in each band there are a licensed PU pair and a SU allowed to transmit by the FC.}

In \eqref{eq:opt_pro}, we in fact maximize the average capacity of a Gaussian channel, where the interference constraint $I_1$ is determined according to the outage constraint on another Gaussian channel $g_1$
\be
\label{eq:Pout}
    \Pro\Bigg( \log\bigg(1+\frac{|g_1|^2 Q_2}{\eta_1 + \underbrace{|h_{11}|^2 P}_{I_1}}\bigg)<R_2 \Bigg) \leq \Pro_{\text{out}},
\ee
where $\eta_1$ is the variance of the Gaussian noise; and $R_2$ is the bit-rate of PU $2$. The outage constraint in \eqref{eq:Pout} yields the interference constraint in \eqref{eq:opt_pro}, given $Q_2,R_2,\eta_1,\Pro_{\text{out}}$. The maximum interference value $I_2$ is written similarly. We assume $I_1$ and $I_2$ are available to SUs. In a careful design, there should be some safety margin between the probability on the left hand-side of \eqref{eq:Pout} and $\Pro_{\text{out}}$ while determining $I_i$. This is because SUs may unintentionally exceed $I_i$ due to lack of information on the true hypothesis and the actual channel coefficients.

\subsection{Spectrum Access Methods}

The conventional spectrum access methods for cognitive radio, namely the opportunistic access and underlay methods, provide simplistic solutions to \eqref{eq:opt_pro}. In particular, the opportunistic access method focuses only on the binary hypothesis test, i.e., spectrum sensing, and conforms to the interference constraints by simply turning off SU Tx, i.e., $P=0$, when $\Hyp_1$ is declared. When $\Hyp_0$ is declared, SU Tx transmits at the maximum power, i.e., $P=P_{\text{max}}$. On the other hand, the underlay method does not perform spectrum sensing and solves only the constrained optimization problem under $\Hyp_1$. As a result, the constant power $P=\min\left\{P_{\text{max}},\frac{I_1}{|h_{11}|^2},\frac{I_2}{|h_{21}|^2}\right\}$ is transmitted under both $\Hyp_0$ and $\Hyp_1$. It is seen that deep fades in the cross links $\{h_{ik}\}$ are beneficial for the SU throughput.

\begin{figure}[t]
\centering
\includegraphics[scale=0.6]{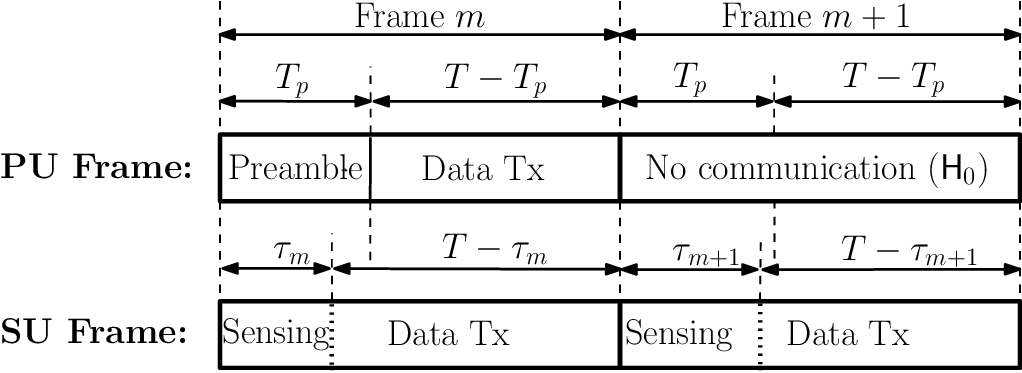}
\caption{Frame structures for the PU communication and the SU communication.}
\label{fig:frame}
\end{figure}

In practice, the channels $g_1,g_2$ between the PUs, and the cross links $\{h_{ik}\}$ are not known a priori. Hence, the PUs perform a preamble communication with duration $T_p$ at the beginning of each data transmission frame to estimate $g_1$ and $g_2$. Specifically, for transmission frame $m$, as shown in Fig. \ref{fig:frame}, PU $i$ estimates $g_i$ during $t\in\big(T(m-1),T(m-1)+T_p\big]$ using pilot symbols, and then data transmission takes place during $t\in\big(T(m-1)+T_p,Tm\big]$, where $T$ is the frame duration. Assuming the SUs are synchronized with the PU frame timing and observe pilot signals, each SU can estimate its cross links during each preamble period.

As opposed to the naive solutions of the conventional spectrum access methods, an efficient solution to \eqref{eq:opt_pro} should involve both spectrum sensing and channel estimation, hence it is a combination of the opportunistic access and underlay methods. For example, at a fixed time $\tau\in(0,T_p]$, we can employ the optimum detector, i.e., the likelihood ratio test (LRT) for spectrum sensing, and the optimum estimator, i.e., the minimum mean square error (MMSE) estimator for channel estimation. Once we obtain the spectrum sensing result ($\Hyp_0$ or $\Hyp_1$) and the channel estimates $\{\hat{h}_{ik}\}$, we can use them to solve \eqref{eq:opt_pro} as follows: SU Tx transmits with $P=P_{\text{max}}$ when $\Hyp_0$ is declared, as in opportunistic access, and with $P=\min\left\{P_{\text{max}},\frac{I_1}{|\hat{h}_{11}|^2},\frac{I_2}{|\hat{h}_{21}|^2}\right\}$ when $\Hyp_1$ is declared, as in underlay.

As a more sophisticated example, instead of performing fixed-sample-size detection and estimation (at a fixed time $\tau$) we can determine the sample number based on the observed samples, resulting in a {\em sequential} method with a random sensing time $\tau$. In particular, we can use the sequential probability ratio test (SPRT) \cite{Wald48}, which is the optimum sequential detector for i.i.d. observations in terms of minimizing the average detection delay, for spectrum sensing, and then use the MMSE estimator at the random sensing time $\tau$ to estimate the unknown channel gains $\{h_{ik}\}$.

However, the above approaches based on separate detection and estimation in general may not yield the optimal solution. In the following section, we propose a new and powerful solution based on sequential joint detection and estimation.

\section{Sequential Joint Spectrum Sensing and Channel Estimation}
\label{sec:sjde}

In this section, we focus on SU Tx, to introduce the sequential joint spectrum sensing and channel estimation framework. Hence, the subscript denoting SU Tx is dropped.

\subsection{Motivation}

In our system model, in each frame $m$ during the preamble period $t\in\big(T(m-1),T(m-1)+T_p\big]$, the signal received by the SU from PU $i$ is given by
\be
\label{eq:signal_model}
  y_i[t] = \left\{ \ba{ll} w_i[t] & \text{if}~~ \Hyp_0 \\ h_i ~p_i[t] + w_i[t] & \text{if}~~ \Hyp_1 \ea \right.,~ i=1,2,~ t=1,2,\ldots,
\ee
where $w_i[t] \sim \cN_c(0,N_0^i)$ is the complex additive white Gaussian noise; $h_i \sim\cN_c(\mu_i,\sigma_i^2)$ is the proper complex channel coefficient between PU $i$ and the SU; and $p_i[t]$ is the complex random pilot signal used in the preamble. The processes $\{p_i[t]\}$ and $\{w_i[t]\}$ are independent and they are independent of the random variable $h_i$.
We assume the SU observes $p_i[t]$ at time $t$, e.g., the SU knows the seed of the random number generator that generates $p_i[t]$, for $i=1,2$.

In \eqref{eq:signal_model}, we would like to decide between $\Hyp_0$ and $\Hyp_1$ as soon as possible, and also estimate $h_i$ if we decide for $\Hyp_1$. In other words, our objective is to have a reliable estimate of the channel coefficient $h_i$ every time we detect the presence of PU communication. Deciding as soon as possible is important because an early sensing time, i.e., small $\tau$, enables the SU to transmit data for a longer period of time, i.e., large $T-\tau$, increasing the SU throughput. On the other hand, the SU transmit power, which is a function of sensing decision and estimates of $\{h_1,h_2\}$, should obey the PU maximum interference constraints. Small $\tau$ may increase the misdetection probability and decrease the estimation accuracy, leading to the violation of such constraints and PU outage. Hence, there is a tradeoff in selecting the $\tau$ value. Conventionally $\tau$ is selected offline, resulting in a fixed-sample-size test. Whereas in a sequential test $\tau$ is determined online, i.e., it depends on the observations, and thus it is random. Although sequential tests are more sophisticated than fixed-sample-size tests, they are much more powerful in minimizing the average sensing time, $\Exp[\tau]$, hence suit better the cognitive radio application.

In the separate detection and estimation approach, the unknown channel gain $h_i$ is treated as a nuisance parameter while performing detection. However, channel estimation is an integral part of the problem of interest. Hence, formulating the problem as a joint detection and estimation problem is a more natural way to obtain better overall performance, i.e., SU throughput. Indeed it was shown in \cite{Yilmaz13_jde} that the combined optimum detector and optimum estimator do not produce the optimum overall detection and estimation performance.

\subsection{Problem Formulation}

Since the results in \cite{Yilmaz13_jde} are obtained for real signals, for analytical convenience in our problem we will treat a complex observation (channel) as two real observations (channels). Specifically, we compute $y_i^1[t] \triangleq \Re(p_i[t]^* y_i[t])$ and $y_i^2[t] \triangleq \Im(p_i[t]^* y_i[t])$, hence instead of \eqref{eq:signal_model} we use the following signal model
\be
\label{eq:signal_model_real}
  y_i^n[t] = \left\{ \ba{ll} w_i^n[t] & \text{if}~~ \Hyp_0 \\ h_i^n ~|p_i[t]|^2 + w_i^n[t] & \text{if}~~ \Hyp_1 \ea \right.,~ i=1,2,~ n=1,2,~ t=1,2,\ldots,
\ee
where $h_i^1 \triangleq \Re(h_i)$, $h_i^2 \triangleq \Im(h_i)$, $w_i^1[t] \triangleq \Re(p_i[t]^* w_i[t])$, and $w_i^2[t] \triangleq \Im(p_i[t]^* w_i[t])$. Note in \eqref{eq:signal_model_real} that $h_i^n\sim\cN\left(\frac{\mu_i}{2},\frac{\sigma_i^2}{2}\right), n=1,2~$; and given $p_i[t]$, the noise $w_i^n[t]\sim\cN\left(0,|p_i[t]|^2 \frac{N_0^i}{2}\right), n=1,2$, and $\{w_i^n[t]\}$ are independent across channels (for different $i$) and time. Similar to \eqref{eq:signal_model}, we want to sequentially decide between $\Hyp_0$ and $\Hyp_1$, and also estimate $h_i^n$ when we decide on $\Hyp_1$. To present the sequential joint detection and estimation (SJDE) problem and the optimum solution to it we first focus on a single channel case, i.e., the signal model in \eqref{eq:signal_model_real} for specific $i,n$ values. In particular, the SU, using its observations $\left\{(y_i^n[t],p_i[t])\right\}_t$ through the real channel $n$ linked to PU $i$, wants to jointly detect the PU communication and estimate the channel coefficient $h_i^n$ when it decides on its presence.

In sequential methods, in general, the average sample number, which corresponds to the average sensing time in our context, is minimized subject to a set of constraints, e.g., false alarm and misdetection constraints for detection, and mean squared error constraint for estimation. In the proposed joint framework we use the following combined cost function
\be
\label{eq:cost}
    \cC\left(\tau,d_{\tau},\hat{x}_{\tau}\right) = c_0 \Pro_0\left(d_{\tau}=1|\cF_{\tau}\right) + c_1 \Pro_1\left(d_{\tau}=0|\cF_{\tau}\right) \\ + c_e \bExp_1\left[\left(\hat{x}_{\tau}-x\right)^2 \ind{d_{\tau}=1} + x^2 \ind{d_{\tau}=0} |\cF_{\tau}\right]
\ee
where $d_{\tau}$ is the decision function; $x \triangleq h_i^n$ is the unknown parameter; $\hat{x}_{\tau}$ is the estimate of $x$; $c_0,c_1,c_e$ are nonnegative constants selected by the designer; $\Pro_0$ and $\Exp_0$ denote the probability measure and expectation under hypothesis $\Hyp_0$; $\bPro_1$ and $\bExp_1$ denote the probability measure and expectation under $\Hyp_1$; $\Pro_1$ and $\Exp_1$ denote the probability measure and expectation under $\Hyp_1$ with $x$ being marginalized; $\cF_t=\sigma\left\{p_i[1],\ldots,p_i[t]\right\}$ is the $\sigma$-algebra, that is, the accumulated history pertinent to the observed process $\left\{p_i[t]\right\}$; and $\ind{A}$ is the indicator of the event $A$, taking the value $1$ if $A$ occurs and $0$ otherwise. Then, our constrained optimization problem is given by
\be
\label{eq:seq_pro}
    \min_{\tau,d_{\tau},\hat{x}_{\tau}} \Exp\left[\tau|\cF_{\tau}\right] ~~\text{subject to}~~ \cC\left(\tau,d_{\tau},\hat{x}_{\tau}\right) \leq \alpha,
\ee
where $\alpha>0$ is a given constant, denoting the target accuracy level.

Our formulation in \eqref{eq:cost} and \eqref{eq:seq_pro} is conditioned on the auxiliary statistic $\cF_t$ because using such extra information we can assess the accuracy of the detector and estimator more precisely than the unconditional formulation. More specifically, since $\Pro_0\left(d_{\tau}=1\right)=\Exp_0\left[\ind{d_{\tau}=1}\right]=
\Exp\left[\Exp_0\left[\ind{d_{\tau}=1}|\cF_{\tau}\right]\right]=
\Exp\left[\Pro_0\left(d_{\tau}=1|\cF_{\tau}\right)\right]$, there is no need to use the expectation, e.g., $\Pro_0\left(d_{\tau}=1\right)$, of an accuracy assessment term when the term itself, e.g., $\Pro_0\left(d_{\tau}=1|\cF_{\tau}\right)$, is available. Moreover, with the conditional formulation used in \eqref{eq:cost} we do not need to specify the distribution of the pilot signal $p_i[t]$.
Note that the constraint $\cC\left(\tau,d_{\tau},\hat{x}_{\tau}\right) \leq \alpha$ in \eqref{eq:seq_pro} is required to hold for each realization of the process $\left\{p_i[t]\right\}$, hence is stricter than its unconditional counterpart, which is required to hold only on average with respect to $\left\{p_i[t]\right\}$.

In \eqref{eq:cost}, the first two terms, which are related to the detection problem, correspond to the false alarm and misdetection probabilities ($\Pro_f$ and $\Pro_m$), respectively. On the other hand, the last term, which is related to the estimation problem, depends on both the decision and estimation strategies. Without this term, i.e., for $c_e=0$, the combined cost depends only on the decision function $d_{\tau}$, implying that the joint problem reduces into a pure detection problem.

Similar to $\cF_t$ let $\cG_t=\sigma\left\{ (y_i^n[1],p_i[1]),\ldots,(y_i^n[t],p_i[t]) \right\}$ denote the $\sigma$-algebra generated by the processes $\{y_i^n[t]\}$ and $\left\{p_i[t]\right\}$, i.e., the complete observation history. Then, we have the corresponding filtrations $\{\cF_t\}_{t\geq0}$ and $\{\cG_t\}_{t\geq0}$. In general, the solution we seek should use all available information, that is, we are looking for a triplet $(\tau,d_{\tau},\hat{x}_{\tau})$ where $\tau$ is $\{\cG_t\}$-adapted, $d_{\tau}$ and $\hat{x}_{\tau}$ are $\cG_{\tau}$-measurable. It is known in the pure estimation problem that with a $\{\cG_t\}$-adapted stopping time $\tau$, in most cases, finding an optimum sequential estimator $(\tau,\hat{x}_{\tau})$ is not tractable \cite{Ghosh87}. Instead, \cite{Grambsch83} considered using an $\{\cF_t\}$-adapted stopping time, which was later shown to have a simple optimal solution for continuous-time and discrete-time observations in \cite{Fellouris12} and \cite{Yilmaz13_est}, respectively. Similarly, in the pure detection problem with a $\{\cG_t\}$-adapted $\tau$ we have a two-dimensional optimal stopping problem, which is not tractable. Consequently, following the approach used for the pure estimation problem in \cite{Grambsch83,Fellouris12,Yilmaz13_est} we consider $\{\cF_t\}$-adapted stopping times for our joint problem. On the other hand, we are still interested in $\cG_{\tau}$-measurable decision rule $d_{\tau}$ and estimator $\hat{x}_{\tau}$, which use all available information acquired up to stopping time $\tau$. As a result, the problem in \eqref{eq:seq_pro} takes the following form
\be
\label{eq:seq_pro_fin}
    \min_{\tau,d_{\tau},\hat{x}_{\tau}} \tau ~~\text{s.t.}~~ \cC\left(\tau,d_{\tau},\hat{x}_{\tau}\right) \leq \alpha.
\ee

\ignore{Assume that the pilot signal $p_i[t]$ has a constant power over time, thus the noise $\{w_{ik}^n[t]\}$ are identically distributed.}

\subsection{The Optimal Solution}

The following theorem gives the optimum solution to the above problem.

\begin{thm}
\label{thm:sjde}
  \ignore{With $|p_i[t]|^2=P$ constant for all $t$, } Consider the observations $\left\{(y_i^n[t],p_i[t])\right\}_t$ obtained through the real channel $x=h_i^n$. Then, the optimum triplet $\left(\tau,d_{\tau},\hat{x}_{\tau}\right)$ of stopping time, decision function, and estimator for the sequential joint detection and estimation (SJDE) problem in \eqref{eq:seq_pro_fin} is given by
  \begin{align}
    \tau =& \min\left\{ t>0: U_t^i\geq\gamma \right\} \label{eq:opt_stop}\\
    d_{\tau} =& \left\{ \begin{array}{ll}
      1 & \text{if}~~ L_{\tau}^{in} \geq \log \frac{c_0}{c_1+c_e \hat{x}_{\tau}^2} \\
      0 & \text{otherwise}
    \end{array} \right. \label{eq:opt_dec}\\
    \hat{x}_t =& \frac{V_t^{in}+\frac{\mu_i}{2}\frac{N_0^i}{\sigma_i^2}} {U_t^i+\frac{N_0^i}{\sigma_i^2}}, \label{eq:opt_est}
  \end{align}
  where $U_t^i \triangleq \sum_{m=1}^t |p_i[m]|^2$ is the conditional Fisher information given $\cF_t$ in estimating $x=h_i^n$ under $\Hyp_1$ [cf. \eqref{eq:signal_model_real}]; $V_t^{in} \triangleq \sum_{m=1}^t y_i^n[m]$; $\gamma$ is a constant threshold \cite[Theorem 1]{Yilmaz13_jde}; and
  \be
  \label{eq:like}
    L_t^{in} \triangleq \frac{\left(V_t^{in}+\frac{\mu_i}{2}\frac{N_0^i}{\sigma_i^2} \right)^2} {N_0^i \left(U_t^i+\frac{N_0^i}{\sigma_i^2}\right)} -\frac{\mu_i^2}{4\sigma_i^2} -\frac{1}{2}\log\left(\frac{\sigma_i^2}{N_0^i}U_t^i+1\right)
  \ee
  is the conditional log-likelihood ratio (LLR) between the hypotheses $\Hyp_0$ and $\Hyp_1$ given $\cF_t$ with $x$ under $\Hyp_1$ being marginalized \cite[Lemma 2]{Yilmaz13_jde}.
\end{thm}

\begin{IEEEproof}
  The proof closely follows \cite{Yilmaz13_jde}, so omitting the details we only highlight the differences here. The main difference is that the noise in \eqref{eq:signal_model_real} is independent across time but has a time-varying variance, whereas i.i.d. noise is assumed in \cite{Yilmaz13_jde}. The common term $|p_i[t]|^2$ in the variance and the mean of the observation $y_i^n[t]$ given $h_i^n$ and $p_i[t]$ under $\Hyp_1$ cancels while writing the estimator $\hat{x}_t$ and the LLR $L_t^{in}$. As a result, the definitions of the Fisher information term $U_t^i$ and its companion $V_t^{in}$ differ from their counterparts in \cite{Yilmaz13_jde}. However, the results in \cite{Yilmaz13_jde} still hold here with the new definitions of $U_t^i, V_t^{in}$ and the noise variance appearing without $|p_i[t]|^2$ as $\frac{N_0^i}{2}$ since $U_t^i, V_t^{in}$ and accordingly other key terms maintain their properties, e.g., $U_t^i$ is increasing.
\end{IEEEproof}
\ignore{
In Theorem \ref{thm:sjde} the optimum triplet is written in its original form as in \cite{Yilmaz13_jde} since we will use it as a reference while developing a distributed spectrum access scheme in the following section. We need to have identically distributed noise over time to borrow the optimal solution from \cite{Yilmaz13_jde}, hence a constant pilot signal power $P$ is assumed. Substituting $|p_i[t]|^2=P$ in \eqref{eq:opt_stop} and \eqref{eq:opt_est} the optimum sensing time and the optimum estimator simplify into $\tau=\lceil \frac{\gamma}{P^2} \rceil$ and $\hat{x}_t=\frac{\sum_{m=1}^t y_{ik}^n[m]+\frac{\mu_{ik}N_0^{ik}}{2\sigma_{ik}^2}} {Pt+\frac{N_0^{ik}}{\sigma_{ik}^2}}$, respectively, where $\lceil\cdot\rceil$ is the ceiling function. Although $\tau=\lceil \frac{\gamma}{P^2} \rceil$ is deterministic with the assumption $|p_i[t]|^2=P$, in general $U_t^i$ and accordingly $\tau$ are random without the assumption.}

The optimum stopping rule in \eqref{eq:opt_stop} terminates getting new samples when the conditional Fisher information exceeds a threshold whose exact expression can be found in \cite[Theorem 1]{Yilmaz13_jde}. Since the conditional Fisher information is increasing, it is guaranteed to have a finite stopping, i.e., sensing, time. The optimum decision function in \eqref{eq:opt_dec} is a modification of the well-known likelihood ratio test (LRT). For $c_e=0$, i.e., in the pure detection problem, it boils down to LRT. For $c_e\not=0$ the estimator is incorporated into LRT. The way it modifies LRT is quite intuitive. When the estimate is nonzero, the threshold is decreased, supporting a decision in favor of $\Hyp_1$. The further the estimate is from zero, the easier to decide for $\Hyp_1$. The estimate provides some side information about the true hypothesis, and the optimum solution to the joint problem uses it. Such a plausible modification appears in the decision function since the detection and estimation problems are formulated jointly. The optimum estimator, given in \eqref{eq:opt_est}, is the minimum mean square error (MMSE) estimator, which is equivalent to the maximum a posteriori (MAP) estimator in the Gaussian case under consideration.

\subsection{Discussions}

Comparing the optimum triplet in Theorem \ref{thm:sjde} with the combined SPRT \& MMSE method, we see that there are fundamental differences in the stopping rule and decision function. In SPRT \cite{Wald47}, the stopping time and detection decision are determined together through a common procedure. More specifically, two thresholds are used to jointly terminate the scheme and make a decision. When the scheme terminates, the decision is already clear as it is determined by the threshold that causes termination. As a result, the performance metrics $\Pro_f$, $\Pro_m$, $\Exp[\tau]$, and also MSE$=\Exp[(\hat{x}_{\tau}-x)^2]$ are closely interrelated since they are all controlled by the two thresholds, which are the only system parameters. On the other hand, in SJDE the stopping time and decision are computed using two separate procedures. First the stopping time is found by performing a single-threshold-test, and then the decision is made via a modified LRT. In particular, $\Exp[\tau]$ and $\Exp[(\hat{x}_{\tau}-x)^2]$ are controlled by only the stopping threshold $\gamma$, whereas $\Pro_f$ and $\Pro_m$ are controlled by $\gamma$, $c_0$, $c_1$, and $c_e$. That is to say, $\Exp[\tau]$ and $\Exp[(\hat{x}_{\tau}-x)^2]$ can be controlled independently from $\Pro_f$ and $\Pro_m$ through $\gamma$, and similarly $\Pro_f$ and $\Pro_m$ can be controlled independently from $\Exp[\tau]$ and $\Exp[(\hat{x}_{\tau}-x)^2]$ through $c_0$, $c_1$, and $c_e$. The latter set of parameters enables a trade-off between $\Pro_f$ and $\Pro_m$ without affecting $\Exp[\tau]$ and $\Exp[(\hat{x}_{\tau}-h_i^n)^2]$. For instance, we can trade false alarm probability $\Pro_f$ for misdetection probability $\Pro_m$, which is crucial for complying with the outage constraints of PUs, by decreasing the ratio of $c_0$ to $c_1$ or $c_e$ without sacrificing early stopping or estimation quality. We obviously have a higher degree of freedom in SJDE than SPRT due to the number of parameters that control the system performance, which endows us with the ability to strike a right balance between our objectives of early stopping, and accurate detection and estimation.

\begin{figure}[t]
\centering
  \includegraphics[width=5in]{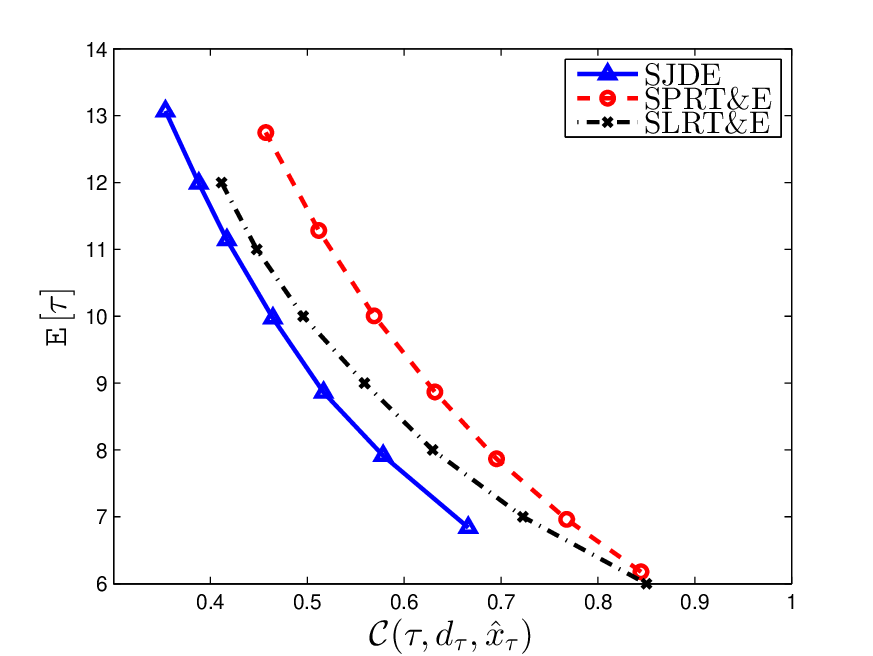}\\
  \caption{Average sensing time vs. combined cost for the SJDE in Theorem \ref{thm:sjde}, and the SPRT \& MMSE and the sequential LRT \& MMSE equipped with the stopping rule of SJDE.}\label{fig:cent}
\end{figure}

In Fig. \ref{fig:cent}, we numerically show the superior performance of SJDE over the combined SPRT \& MMSE (SPRT\&E) in terms of the combined detection and estimation cost in \eqref{eq:cost}. We also compare SJDE with the sequential LRT \& MMSE (SLRT\&E) that is equipped with the stopping rule of SJDE to demonstrate the advantage of incorporating the estimate into the decision function. SLRT\&E uses the unmodified (original) LRT to detect, hence can be seen as a separate-formulation-method. It outperforms SPRT\&E since it enjoys the flexibility of SJDE to strike a desired balance for the specific problem of interest by employing two separate procedures, namely the stopping rule of SJDE and LRT, to terminate the scheme and make a decision respectively. In our problem of interest, it is crucial that SUs do not violate the maximum interference constraint, which in turn ensures an admissible PU outage probability. In case of misdetection the SU transmits with maximum power, which may cause the violation of outage constraint. Even when the SU correctly detects PU communication, poor channel estimate may still cause the SU to transmit with a non-admissible power. On the other hand, the false alarm, which corresponds to deciding on $\Hyp_1$ under $\Hyp_0$, is not related to the outage constraint, but only degrades the SU throughput. Therefore, in the combined cost expression in \eqref{eq:cost} the second and third terms are more important than the first term. Accordingly, in Fig. \ref{fig:cent} we use $c_0=c_1=0.2$ and $c_e=0.6$. Since the second part of the third term in \eqref{eq:cost} already penalizes misdetection, we do not differentiate between the coefficients, $c_0$ and $c_1$, of the detection error probabilities. In Fig. \ref{fig:cent}, referring to \eqref{eq:signal_model_real} we use $\mu_i=0$, i.e., Rayleigh fading channel $h_i^n$, and $\sigma_i^2=N_0^i=\Exp[|p_i[t]|^2]=1$.

\subsection{SJDE for a Single SU with Multiple Channels}

Here, following the optimum SJDE scheme in Theorem \ref{thm:sjde} for the single channel case we are interested in finding the optimum SJDE scheme for the SU observing the signals $\left\{y_i^n[t]\right\}$ and $\left\{p_i[t]\right\}$ through the channels $\left\{h_i^n\right\},~i=1,2,~n=1,2$ from PU $1$ and PU $2$. We first need to modify the cost function in \eqref{eq:cost} by adding the new MSE terms, i.e.,
\begin{multline}
\label{eq:cost_mult}
    \cC\left(\tau,d_{\tau},\hat{h}_i^n[\tau]\right) = c_0 \Pro_0\left(d_{\tau}=1|\cF_{\tau}\right) + c_1 \Pro_1\left(d_{\tau}=0|\cF_{\tau}\right) \\ + c_e \sum_{i=1}^2 \sum_{n=1}^2 \bExp_1\left[\left(\hat{h}_i^n[\tau]-h_i^n\right)^2 \ind{d_{\tau}=1} + (h_i^n)^2 \ind{d_{\tau}=0} \Big|\cF_{\tau}\right].
\end{multline}
The following theorem, whose proof is provided in the Appendix, gives the optimum SJDE scheme in this case.
\begin{thm}
\label{thm:mult}
  With the cost function in \eqref{eq:cost_mult}, and the observations $\left\{y_i^n[t],p_i[t]\right\}$ obtained through the channels $\left\{h_i^n\right\},~i=1,2,~n=1,2$ from PU $1$ and PU $2$, the optimum triplet of stopping time, decision function, and estimator for the sequential joint
  detection and estimation (SJDE) problem in \eqref{eq:seq_pro_fin} is given by
    \begin{align}
    \tau =& \min\left\{ t>0: U_t\geq\bar{\gamma} \right\} \label{eq:opt_stop_SU}\\
    d_{\tau} =& \left\{ \begin{array}{ll}
      1 & \text{if}~~ L_{\tau} \geq \log \frac{c_0}{c_1+c_e \sum_{i=1}^2 \sum_{n=1}^2 \left(\hat{h}_i^n[\tau]\right)^2} \\
      0 & \text{otherwise}
    \end{array} \right. \label{eq:opt_dec_SU}\\
    \hat{h}_i^n[t] =& \frac{V_t^{in}+\frac{\mu_i}{2}\frac{N_0^i}{\sigma_i^2}} {U_t^i+\frac{N_0^i}{\sigma_i^2}}, ~~\forall i,n, \label{eq:opt_est_SU}
  \end{align}    
  where $U_t=\sum_{i=1}^2 \sum_{m=1}^t |p_i[m]|^2$ is the conditional Fisher information given $\cF_t$ under $\Hyp_1$; $\bar{\gamma}$ is a constant threshold [cf. \eqref{eq:opt_stop}]; and $L_t=\sum_{i=1}^2 \sum_{n=1}^2 L_t^{in}$ [cf. \eqref{eq:like}] is the global LLR. 
\end{thm}

For systems with multiple SU pairs, in the next section we propose a distributed and cooperative spectrum access method which selects the SU with the maximum achievable throughput, and controls its transmit power.

\section{Distributed Spectrum Access Based on SJDE}
\label{sec:dsa}

In the previous section we formulated the joint spectrum sensing and channel estimation problem for a single SU and gave the optimal solution to it. In this section we consider $K$ SUs, i.e., $K/2$ SU transmitter-receiver pairs, where each SU observes signals through $4$ different real channels ($2$ from each PU). All observations of $K$ SUs through $4K$ channels are used to detect a single event, namely the PU communication. Hence, under the joint framework introduced in Section \ref{sec:sjde}, SUs can cooperate to detect the PU communication. We next propose a bandwidth and energy-efficient distributed spectrum access algorithm for the cognitive radio system under consideration.

\subsection{SJDE-based Spectrum Access with Multiple SUs}
\label{sec:levtrig}

We now consider the multi-SU case for SJDE, and propose a dynamic spectrum access method (DSA-SJDE). 
From \eqref{eq:cost_mult}, we have the following cost function,
\begin{multline}
\label{eq:cost_mult_K}
    \cC\left(\tau,d_{\tau},\hat{h}_i^n[\tau]\right) = c_0 \Pro_0\left(d_{\tau}=1|\cF_{\tau}\right) + c_1 \Pro_1\left(d_{\tau}=0|\cF_{\tau}\right) \\ + c_e \sum_{k=1}^K \sum_{i=1}^2 \sum_{n=1}^2 \bExp_1\left[\left(\hat{h}_{ik}^n[\tau]-h_{ik}^n\right)^2 \ind{d_{\tau}=1} + (h_{ik}^n)^2 \ind{d_{\tau}=0} \Big|\cF_{\tau}\right].
\end{multline}
Note that all $K$ SUs observe the same pilot signals $\left\{p_1[t]\right\}$ and $\left\{p_2[t]\right\}$. Hence, from \eqref{eq:opt_stop_SU} it is seen that they have the same stopping time, which in this case serves as a global stopping time. Each channel coefficient $h_{ik}^n$ is again estimated using \eqref{eq:opt_est_SU} for all $k,i,n$ because they are independent. Since the observations $\left\{y_{ik}^n[t]\right\}_k$, across SUs, are independent given $\left\{p_i[t]\right\}$, the global LLR is written as $L_t=\sum_{k=1}^K L_t^k$, and as in \eqref{eq:opt_dec_SU} we sum the channel estimates to write the threshold. Then, substituting the global LLR and the global threshold in \eqref{eq:opt_dec_SU} we obtain the decision function for the multi-SU case.

\begin{cor}
In the multi-SU case with the cost function in \eqref{eq:cost_mult_K}, the optimum solution to \eqref{eq:seq_pro_fin} is given  by
\begin{align}
    \tau =& \min\left\{ t>0: U_t\geq\bar{\bar{\gamma}} \right\} \label{eq:opt_stop_SU_K}\\
    d_{\tau} =& \left\{ \begin{array}{ll}
      1 & \text{if}~~ L_{\tau} \geq \log \frac{c_0}{c_1+c_e \sum_{k=1}^K \sum_{i=1}^2 \sum_{n=1}^2 \left(\hat{h}_{ik}^n[\tau]\right)^2} \\
      0 & \text{otherwise}
    \end{array} \right. \label{eq:opt_dec_SU_K}\\
    \hat{h}_{ik}^n[t] =& \frac{V_t^{ikn}+\frac{\mu_{ik}}{2}\frac{N_0^{ik}}{\sigma_{ik}^2}} {U_t^i+\frac{N_0^{ik}}{\sigma_{ik}^2}}, ~~\forall i,k,n, \label{eq:opt_est_SU_K}
  \end{align}    
  where $U_t=\sum_{i=1}^2 \sum_{m=1}^t |p_i[m]|^2$ is the conditional Fisher information given $\cF_t$ under $\Hyp_1$; $\bar{\bar{\gamma}}$ is a constant threshold [cf. \eqref{eq:opt_stop}]; $V_t^{ikn} = \sum_{m=1}^t y_{ik}^n[m]$ [cf. \eqref{eq:signal_model_real}]; and $L_t=\sum_{k=1}^K \sum_{i=1}^2 \sum_{n=1}^2 L_t^{ikn}$ [cf. \eqref{eq:like}] is the global LLR. 
\end{cor}

It looks like the SJDE scheme for the multi-SU case simply follows from \eqref{eq:opt_stop_SU}--\eqref{eq:opt_est_SU} in the single-SU case. However, in the multi-SU case the stopping time, detector, and estimator are computed at the FC, which requires some local information. Note that the FC can reasonably observe the pilot signals $\left\{p_1[t]\right\}$ and $\left\{p_2[t]\right\}$ in the same way SUs do. Then, the FC needs to know the local random variables $\left\{V_{\tau}^{ikn}\right\}_{i,k,n}$ at the stopping time $\tau$. In a straightforward way SUs can quantize and send $\left\{V_{\tau}^{ikn}\right\}_{i,k,n}$ at time $\tau$. However, this method has several disadvantages in practice. Firstly, it needs high bandwidth at time $\tau$ on each reporting channel between SUs and the FC. Moreover, the reporting channels are utilized inefficiently. They remain idle until time $\tau$, and at time $\tau$ each SU sends a number of bits, which may cause congestion at the FC. To overcome these practical issues SUs can sequentially report $\left\{V_{\tau}^{ikn}\right\}_{i,k,n}$. For sequential reporting level-triggered sampling, a non-uniform sampling technique, was shown to be much superior to the traditional uniform sampling in terms of bandwidth and energy requirements  for detection and estimation purposes in \cite{Yilmaz12} and \cite{Yilmaz13_est}, respectively. Therefore, we propose that SUs sequentially report $\left\{V_{\tau}^{ikn}\right\}_{i,k,n}$ using level-triggered sampling.

\subsection{Level-triggered Sampling}
\begin{figure}[t]
\centering
  \includegraphics[width=4.5in]{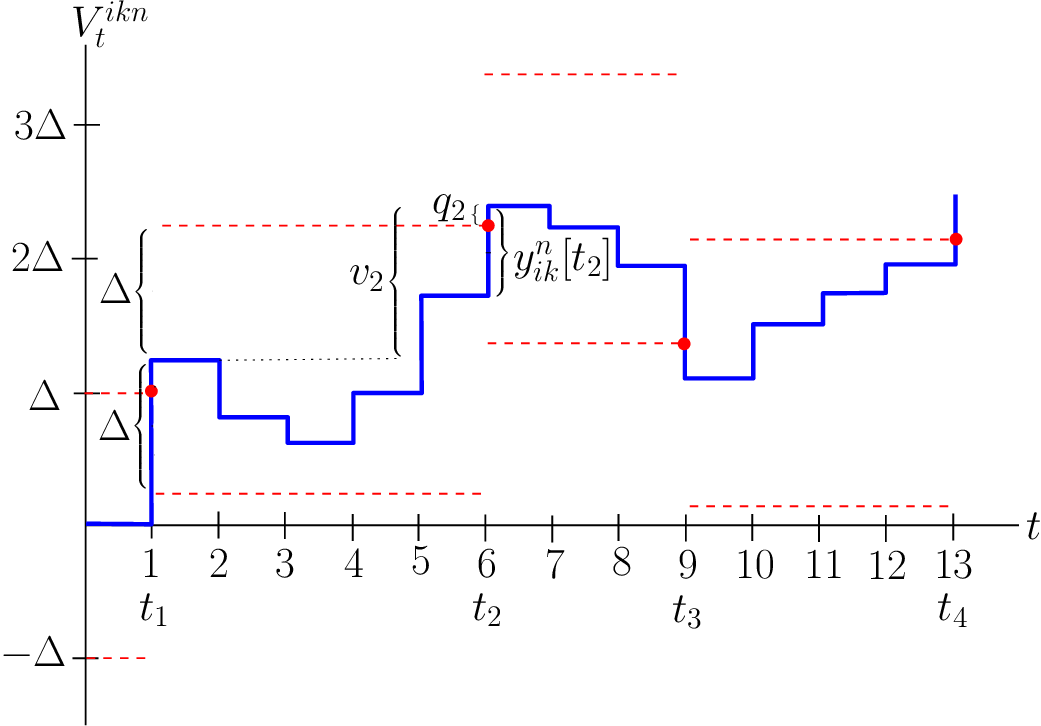}\\
  \caption{The level-triggered sampling procedure used at SUs.}\label{fig:levtrig}
\end{figure}

Each SU $k$, via the same level-triggered sampling procedure, informs the FC whenever considerable change occurs in its four local processes $\left\{V_t^{ikn}\right\},~i=1,2,~n=1,2$. In other words, $4K$ identical samplers run in parallel for $4K$ different processes. Hence, we will describe the procedure for a single process $\left\{V_t^{ikn}\right\}_t$.
The level-triggered sampling is a simple form of event-triggered sampling, in which sampling (communication) times $\{t_m\},~m\in\bN,$ are not deterministic, but rather dynamically determined by the random process $\left\{V_t^{ikn}\right\}_t$, i.e.,
\be
    \label{eq:samp}
    t_m \triangleq \min\{t>t_{m-1} : V_t^{ikn}-V_{t_{m-1}}^{ikn} \not\in (-\Delta,\Delta)\},~m\in\bN,~t_0=0.
\ee
The threshold parameter $\Delta$ is a constant known by both SUs and the FC.
At each sampling time $t_m$, SU $k$ transmits $r$ bits, $b_{m,1} b_{m,2} \ldots b_{m,r}$, to the FC. The first bit, $b_{m,1}$, indicates the threshold crossed (either $\Delta$ or $-\Delta$) by the incremental process $v_m \triangleq V_{t_m}^{ikn}-V_{t_{m-1}}^{ikn}$, i.e.,
\be
    \label{eq:sign}
    b_{m,1}=\text{sign}(v_m).
\ee
The remaining $r-1$ bits are used to quantize the over(under)shoot $q_m \triangleq |v_m|-\Delta$ into $\tilde{q}_m$. At each sampling time $t_m$, the overshoot value $q_m$ cannot exceed the magnitude of the last sample $|y_{ik}^n[t_m]|$ in the incremental process $v_m=\sum_{t=t_{m-1}+1}^{t_m} y_{ik}^n[t]$. The quantization interval $[0,\phi]$ is uniformly divided into $2^{r-1}$ subintervals with the step size $\frac{\phi}{2^{r-1}}$. The mid value of each subinterval is used as the corresponding quantization level, i.e., a mid-riser quantizer is used. When $q_m>\phi$, the uppermost quantization level is used. The parameter $\phi$ is determined so that $\Pro(q_m>\phi)$ is sufficiently small. From \cite[Section IV-B]{Yilmaz12} we can set the threshold $\Delta$ using
\be
\label{eq:Delta}
    \Delta \tanh\left(\frac{\Delta}{2}\right)=\frac{1}{M}\sum_{k=1}^K\sum_{i=1}^2\sum_{n=1}^2 |\Exp_i[V_1^{ikn}]|
\ee
for the FC to receive messages with an average rate of $M$ messages per unit time under $\Hyp_i,~i=0,1$. In Fig. \ref{fig:levtrig}, the level-triggered sampling procedure is demonstrated on a sample path of $V_t^{ikn}$.

The FC, upon receiving the bits $b_{m,1} b_{m,2} \ldots b_{m,r}$ from SU $k$ at time $t_m$, recovers the quantized value of $v_m$ by computing
\be
    \label{eq:VincrRecov}
    \tilde{v}_m \triangleq b_{m,1}(\Delta+\tilde{q}_m).
\ee
Then, it sequentially sums up $\{\tilde{v}_m\}$, at the sampling (communication) times $\{t_m\}$ to obtain an approximation $\tilde{V}_t^{ikn}$ to the sufficient statistic $V_t^{ikn}$, i.e.,
\be
    \label{eq:VRecov}
    \tilde{V}_t^{ikn} \triangleq \sum_{m=1}^{M_t} \tilde{v}_m,
\ee
where $M_t$ is the number of messages that the FC receives from SU $k$ about the process $\{V_t^{ikn}\}$ up to time $t$.
During the times the FC receives no message, i.e., $t \not\in \{t_m\}$, $\tilde{V}_t^{ikn}$ is kept constant.

At the stopping time $\tau$, given by \eqref{eq:opt_stop_SU}, the FC estimates each channel coefficient $h_{ik}^n$ using
\be
\label{eq:FC_est}
    \tilde{h}_{ik}^n[\tau] = \frac{\tilde{V}_t^{ikn}+\frac{\mu_{ik}}{2}\frac{N_0^{ik}}{\sigma_{ik}^2}} {U_t^i+\frac{N_0^{ik}}{\sigma_{ik}^2}},
\ee
and decides according to the following rule
\be
\label{eq:FC_dec}
    \tilde{d}_{\tau} = \left\{ \begin{array}{ll}
      1 & \text{if}~~ \tilde{L}_{\tau} \geq \log \frac{c_0}{c_1+c_e \sum_{k=1}^K \sum_{i=1}^2 \sum_{n=1}^2 \left(\tilde{h}_{ik}^n[\tau]\right)^2} \\
      0 & \text{otherwise}
    \end{array} \right.,
\ee
where $\tilde{L}_{\tau}=\sum_{k=1}^K \sum_{i=1}^2 \sum_{n=1}^2 \tilde{L}_{\tau}^{ikn}$, and $\tilde{L}_{\tau}^{ikn}$ is computed from \eqref{eq:like} by substituting $\tilde{V}_t^{ikn}$ for $V_t^{ikn}$. After making a decision, the FC grants the transmission privilege to the SU Tx with the highest achievable throughput. When the decision is in favor of $\Hyp_0$, i.e., $\tilde{d}_{\tau}=0$, one of them is selected randomly (or in some specific order) since in this case any SU Tx can transmit with its maximum power $P_{\text{max}}$. On the other hand, when $\tilde{d}_{\tau}=1$, the FC selects SU Tx $k^*$ where
\be
\label{eq:kstar}
    k^*\triangleq \arg\max_{k_t} \left\{ \min\left\{ \frac{I_1}{|\tilde{h}_{1k_t}[\tau]|^2},\frac{I_2}{|\tilde{h}_{2k_t}[\tau]|^2} \right\} \right\},
\ee
$k_t$ is the SU Tx index, and $|\tilde{h}_{ik}[\tau]|^2=\sum_{n=1}^2\left(\tilde{h}_{ik}^n[\tau]\right)^2,~i=1,2$.
The pseudocodes for the procedures at SU $k$ and the FC in the proposed SJDE-based dynamic spectrum access method (DSA-SJDE) are given in Algorithms \ref{alg:SU} and \ref{alg:FC}, respectively. In Algorithm \ref{alg:SU}, an SU Rx never executes lines 17-23 since the FC reports $d_{\tau}$ to SU Tx $k^*$ (cf. line 20 in Algorithm \ref{alg:FC}).

\begin{algorithm}[h!]\small
\caption{\small DSA-SJDE procedure at SU $k$}
\label{alg:SU}
\baselineskip=0.5cm
\begin{algorithmic}[1]
\STATE Initialization: $\left\{t,m_{in},v_{in},V_{in},U_i\right\} \gets 0, ~\forall i,n$
\WHILE {$|v_{in}| < \Delta, \forall i,n$ \AND $\sum_{i=1}^2 U_i<\gamma$}
    \STATE $t \gets t+1$
    \STATE $v_{in} \gets v_{in} + y_{ik}^n[t]$
    \STATE $V_{in} \gets V_{in} + y_{ik}^n[t]$
    \STATE $U_i \gets U_i + |p_i[t]|^2$
\ENDWHILE
\IF {$|v_{in}| \geq \Delta$ \COMMENT{for any $i,n$}}
    \STATE $m_{in} \gets m_{in}+1$
    \STATE $t_m^{in}=t$
    \STATE Send $b_{m,1}^{in}=\text{sign}(v_{in})$ and $r-1$ quantization bits for $q_m^{in}=v_{in}-\Delta$ to FC
    \STATE $v_{in} \gets 0$
\ENDIF
\IF {$\sum_{i=1}^2 U_i\geq\gamma$ \OR $t\geq T_p$}
    \STATE $\tau=t$
    \IF {FC reports $d_{\tau}$}
        \IF {$d_{\tau}=0$}
            \STATE $P=P_{\text{max}}$
        \ELSE
            \STATE Compute $\hat{h}_{in}$ as in \eqref{eq:opt_est_SU} using $V_{in}$ and $U_i$
            \STATE $P=\min\left\{P_{\text{max}},\frac{\hat{I}_1^{\tau}}{\sum_{n=1}^2 (\hat{h}_{1n})^2},\frac{\hat{I}_2^{\tau}}{\sum_{n=1}^2 (\hat{h}_{2n})^2}\right\}$,~ \COMMENT{see \eqref{eq:Ihat} for $\hat{I}_i^{\tau}$}
        \ENDIF
        \STATE Start data transmission with power $P$
    \ELSE
        \STATE Stop
    \ENDIF
\ELSE
    \STATE Go to line 2
\ENDIF
\end{algorithmic}
\end{algorithm}

\begin{algorithm}[h!]\small
\caption{\small DSA-SJDE procedure at FC}
\label{alg:FC}
\baselineskip=0.5cm
\begin{algorithmic}[1]
\STATE Initialization: $\left\{V_{ikn},U_i\right\} \gets 0, ~\forall i,k,n$
\WHILE {$\sum_{i=1}^2 U_i<\gamma$ \OR $t<T_p$}
    \STATE $t \gets t+1$
    \STATE $U_i \gets U_i + |p_i[t]|^2$
    \IF {$b_{m,1}^{ikn}\ldots b_{m,r}^{ikn}$ received \COMMENT{for any $i,k,n$}}
        \STATE Compute $\tilde{q_m^{ikn}}$ from $b_{m,2}^{ikn}\ldots b_{m,r}^{ikn}$
        \STATE $V_{ikn}=V_{ikn}+b_{m,1}^{ikn}(\Delta+\tilde{q_m^{ikn}})$
    \ENDIF
\ENDWHILE
\STATE $\tau=t$
\STATE Compute $\tilde{h}_{ikn}$  from \eqref{eq:FC_est} using $V_{ikn}$ and $U_i,~\forall i,k,n$

\STATE Compute $L_{ikn}$ from \eqref{eq:like} using $V_{ikn}$ and $U_i,~\forall i,k,n$
\STATE $L=\sum_{k=1}^K \sum_{i=1}^2 \sum_{n=1}^2 L_{ikn}$
\STATE Compute $d_{\tau}$ from \eqref{eq:FC_dec} using $L$ and $\{\tilde{h}_{ikn}\}$
\IF {$d_{\tau}=0$}
    \STATE Select $k^*$ randomly or in some specific order from SU transmitters
\ELSE
    \STATE Find $k^*$ as in \eqref{eq:kstar} using $\{\tilde{h}_{ikn}\}$
\ENDIF
\STATE Report $d_{\tau}$ to SU Tx $k^*$, and instruct the others to stop
\end{algorithmic}
\end{algorithm}

\subsection{Discussions}

The procedures at SUs and the FC, given in Algorithms \ref{alg:SU} and \ref{alg:FC}, restarts at the beginning of each frame with duration $T$ (see Fig. \ref{fig:frame}). Each SU $k$ performs the procedure in Algorithm \ref{alg:SU}.  The stopping threshold $\gamma$ is selected through offline simulations to maximize the average SU throughput in DSA-SJDE, given by
\be
\label{eq:avg_thr}
	\bar{R}=\Exp\left[ \frac{T-\tau}{T} \Big\{\big[\pi_0(1-\Pro_f)+(1-\pi_0)\Pro_m\big]\Gamma_0 + \big[\pi_0\Pro_f+(1-\pi_0)(1-\Pro_m)\big]\Gamma_1 \Big\} \right]
\ee
where $\Gamma_0\triangleq\log\left(1+\frac{|\beta_{k^*}|^2 P_{\text{max}}}{N_0^{k^*}}\right)$, $\Gamma_1\triangleq\log\left(1+\frac{|\beta_{k^*}|^2 P_{k^*}^{\tau}}{N_0^{k^*} + |h_{1k^*_r}|^2 Q_1 + |h_{2k^*_r}|^2 Q_2}\right)$, $k^*_r$ denotes the SU Rx corresponding to SU Tx $k^*$, $\pi_0$ is the prior probability for the hypothesis $\Hyp_0$, $\Pro_f$ is the false alarm probability, i.e., $\Pro_0(\tilde{d}_{\tau}=1)$, and $\Pro_m$ is the misdetection probability, i.e., $\Pro_1(\tilde{d}_{\tau}=0)$. The sensing time $\tau$ is governed by the threshold $\gamma$. The scaling term $\frac{T-\tau}{T}$ in \eqref{eq:avg_thr} represents the throughput penalty due to sensing. Hence, small threshold $\gamma$ on average increases the scaling term, affecting $\bar{R}$ positively. On the other hand, it causes larger error probabilities, $\Pro_f$ and $\Pro_m$. Note that $P_{k^*}^{\tau}=\min\left\{ P_{\text{max}}, \frac{\hat{I}_{1k^*}^{\tau}}{|\hat{h}_{1k^*}[\tau]|^2}, \frac{\hat{I}_{2k^*}^{\tau}}{|\hat{h}_{2k^*}[\tau]|^2} \right\} \leq P_{\text{max}}$, thus $\Gamma_0>\Gamma_1$. As a result, increasing $\Pro_f$ decreases $\bar{R}$. Although it looks like $\bar{R}$ is directly proportional to $\Pro_m$, large $\Pro_m$ values are not feasible due to the interference constraints. This defines a lower bound on the stopping threshold $\gamma$. As clearly seen, there is a trade-off in selecting the $\gamma$ value. It is convenient to find the best $\gamma$ value, that maximizes $\bar{R}$, performing an offline numerical search in the interval $[\gamma_0,\gamma_1]$. The lower bound $\gamma_0$ is determined by the interference constraints as mentioned earlier. We need the upper bound $\gamma_1$ to control the probability that the sensing time exceeds the preamble duration, i.e., $\Pro(\tau>T_p)$, where the signal model in \eqref{eq:signal_model_real} is valid. In such an exceptional case, when $\tau>T_p$, the sensing and estimation should terminate, i.e., $\tau=T_p$, since the signal model is no more valid.

When PU communication is detected, i.e., $\tilde{d}_{\tau}=1$, the SU selected for data transmission needs to use calibrated maximum interference levels $\hat{I}_{ik^*}^{\tau}\triangleq\alpha_{ik^*}^{\tau}I_i$, instead of original values $I_i,~i=1,2$, in computing its transmission power. This is required to compensate for estimation errors. To satisfy the interference constraints we should have
\be
\label{eq:Ihat}
    \frac{\alpha_{ik^*}^{\tau}I_i}{|\hat{h}_{ik^*}[\tau]|^2} |h_{ik^*}|^2 \leq I_i,~i=1,2,
\ee
hence $\alpha_{ik^*}^{\tau}\leq\frac{|\hat{h}_{ik^*}[\tau]|^2}{|h_{ik^*}|^2}$ with a high probability. Since the actual channel coefficient $h_{ik^*}$ is unknown, through offline simulations we set $\alpha_{ik_t}^{\tau}$ for each $\tau\in(0,T_p]$ so that $\Pro\left(\frac{|\hat{h}_{ik_t}[\tau]|^2}{|h_{ik_t}|^2}\geq\alpha_{ik_t}^{\tau}\right)$ is sufficiently high. Note that there are two sources that cause excess interference over $I_i$, namely misdetection and the event $\frac{|\hat{h}_{ik^*}[\tau]|^2}{|h_{ik^*}|^2}<\alpha_{ik^*}^{\tau}$. The probabilities $\Pro_m=\Pro_1(\tilde{d}_{\tau}=0)$ and $\Pro\left(\frac{|\hat{h}_{ik^*}[\tau]|^2}{|h_{ik^*}|^2}<\alpha_{ik^*}^{\tau}\right)$ should be made sufficiently small in order to meet the PU outage constraints.

\section{Simulation Results}
\label{sec:sim}

In this section, we provide simulation results to compare different spectrum access methods in terms of the average SU throughput. We first consider two conventional methods: underlay and opportunistic access. These two methods have intrinsic deficiencies. In the former the SU is blind to the idle state of PUs, and in the latter it is unable to benefit from deep fades in cross links. It could be anticipated that a combination of these two methods, as in DSA-SJDE and DSA-SPRT, may result in a higher SU throughput. DSA-SPRT is the straightforward sequential implementation of such combination. It uses SPRT for spectrum sensing, MMSE estimator for channel estimation, and uniform sampling for distributed operation. On the other hand, DSA-SJDE, the proposed novel spectrum access method, uses the SJDE for sensing and estimation, and level-triggered sampling for distributed implementation. In the opportunistic access scheme, we use the LRT for sensing and the traditional uniform sampling for distributed implementation. In the underlay scheme, we assume that SUs somehow perfectly estimate the channel coefficients during the preamble.

We plot the average SU throughput $\bar{R}$ against the outage probability constraint $\Pro_{\text{out}}$, the maximum transmission power $P_{\text{max}}$ for SU, the prior probability $\pi_0$ of idle PU, and the fraction $\frac{T}{T_P}$ of frame length to the preamble duration respectively in the subsequent figures. The preamble duration is fixed at $T_p=10$ ms and the global clock runs, i.e., PUs transmit pilot symbols and SUs observe discrete-time samples, with a frequency of $f_s=1$ MHz. In PU communication 16-QAM is used with an average power $\Exp[|p_i[t]|^2]=P_i=1$. PUs utilize random number generators, whose seeds are known to SUs and the FC, to generate pilot symbols in the preamble. All simulated channels are Rayleigh fading channels, i.e., channel coefficient $h_{ik}$ is proper complex Gaussian random variable with zero mean and finite variance $\sigma_{ik}^2$. We set $N_0^{ik}=\sigma_{ik}^2=1$, hence SNR$=\Exp[|p_i[t]|^2]=1$ ($0$ dB) under $\Hyp_1$.
In opportunistic access and DSA-SPRT, the period of uniform sampling for reporting $V_t^{ikn}$ is set as four unit time, i.e., $T_u=4T_s=\frac{4}{f_s}$. Since each SU samples four processes, the FC receives $K$ messages per unit time, $T_s$. For a fair comparison we set the average message rate of level-triggered sampling to the same value, i.e., $M=K$. Then, using \eqref{eq:Delta} the corresponding value of the sampling threshold $\Delta$ is found. Throughout this section we simulate a two-SU system, i.e., $K=2$.

We use a 50\% safety margin while determining the maximum interference level $I_i$ from $\Pro_{\text{out}}$ using \eqref{eq:Pout}. Moreover, as additional safety measures to protect the PU QoS, i.e., to satisfy the $\Pro_{\text{out}}$ constraint, we determine $\alpha_{i1}^{\tau}$ as the fifth percentile of $\frac{|\hat{h}_{i1}[\tau]|^2}{|h_{i1}|^2}$ to calibrate the maximum interference levels at SUs, and confine the misdetection probability $\Pro_m$ to values smaller than $\Pro_{\text{out}}/5$. For the DSA-SPRT, DSA-SJDE, and the opportunistic access scheme, through offline simulations we find the best parameters that maximize $\bar{R}$, complying with the constraint $\Pro_m<\Pro_{\text{out}}/5$. Specifically, via offline numerical search, we use the optimum values for the threshold pair in SPRT, the stopping threshold $\gamma$ in SJDE, the deterministic sensing time $\tau$ and the LRT threshold in the opportunistic access scheme. We use $c_0=c_1=0.2,~c_e=0.6$ for SJDE as in Section \ref{sec:sjde}.

\underline{{\bf SU throughput vs. PU outage probability}}: In the first set of simulations, we set $P_{\text{max}}=15$ dB, $\pi_0=0.5$, $T=10\times{T_p}$, and vary $\Pro_{\text{out}}\in[0.025,0.125]$. In this case, the maximum interference levels $I_i$ vary between $-9$ dB and $6$ dB.

\begin{figure}[t]
\centering
\includegraphics[width=.8\linewidth]{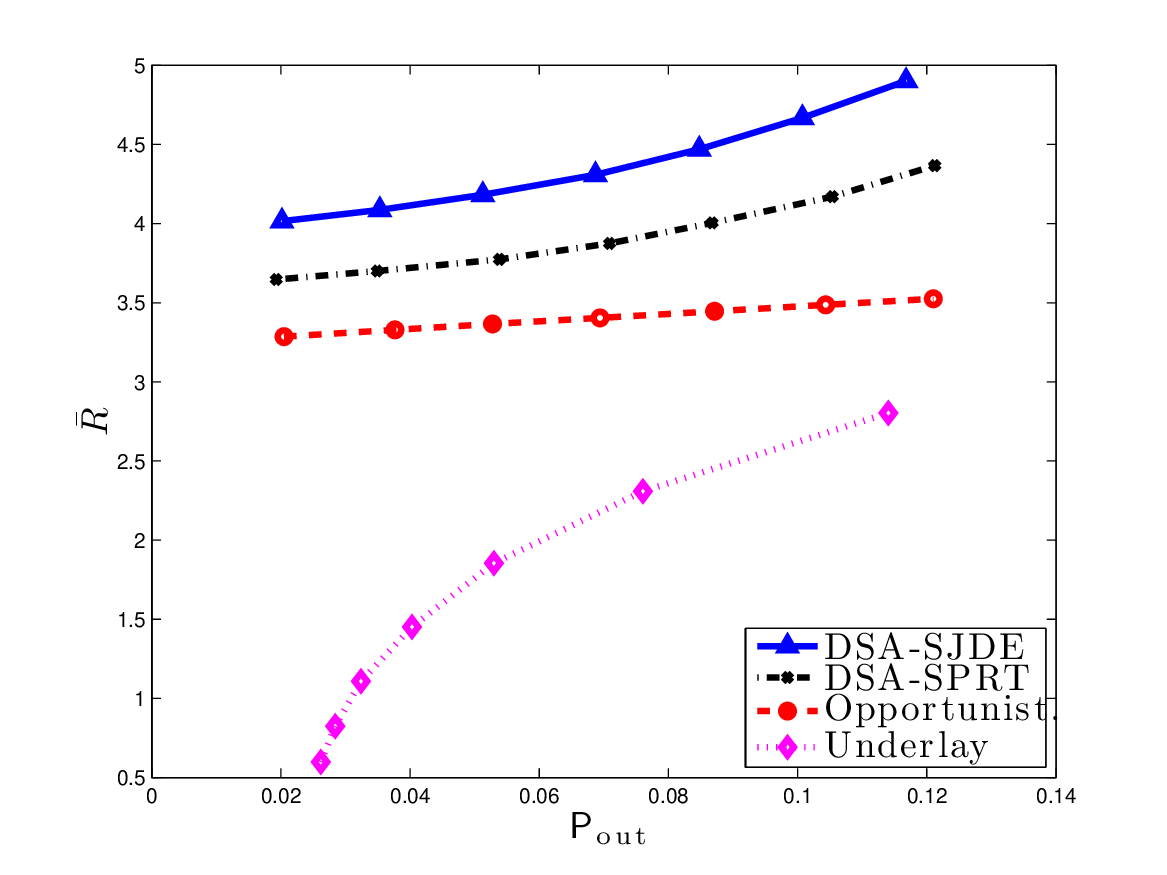}
\caption{Average SU throughput vs. PU outage probability for the conventional (underlay and opportunistic) and the proposed (DSA-SJDE and DSA-SPRT) dynamic spectrum access schemes.}
\label{fig:R_Pout}
\end{figure}

In Fig. \ref{fig:R_Pout}, we see that the proposed spectrum access schemes with sequential detectors and estimators, being combinations of conventional methods, perform better than the underlay and the opportunistic access schemes, as expected. Not surprisingly, the underlay scheme performs poorly under strict outage probability (interference) constraints, and considerably improves its performance as the constraints relax because its transmit power solely depends on the maximum interference levels. Conversely, the opportunistic access scheme is mostly unaffected by the changing outage probability constraint as it does not utilize the maximum interference levels to determine its transmit power. The slight performance increase as $\Pro_{\text{out}}$ grows is due to the relaxation on the $\Pro_m$ constraint.
On the other hand, the sequential schemes, being combinations of the conventional approaches, enjoy the advantages of opportunistic access and underlay when $\Pro_{\text{out}}$ is small and large, respectively. Moreover, the novel DSA-SJDE scheme significantly outperforms DSA-SPRT, which uses well-known techniques for sampling and distributed implementation, due to its distinct features: the joint nature of detector and estimator (cf. Section \ref{sec:sjde}), the separation property of stopping rule and detector (cf. Section \ref{sec:sjde}), and the adaptive nature of level-triggered sampling (cf. Section \ref{sec:levtrig}). Note that the estimator provides some side information about the true hypothesis, and thus its incorporation into the decision function improves the SU throughput, which is a joint function of detector and estimator. For the advantages of the latter two features we refer to Section \ref{sec:sjde} and Section \ref{sec:levtrig}, respectively.

\begin{figure}[t]
\centering
\includegraphics[width=.8\linewidth]{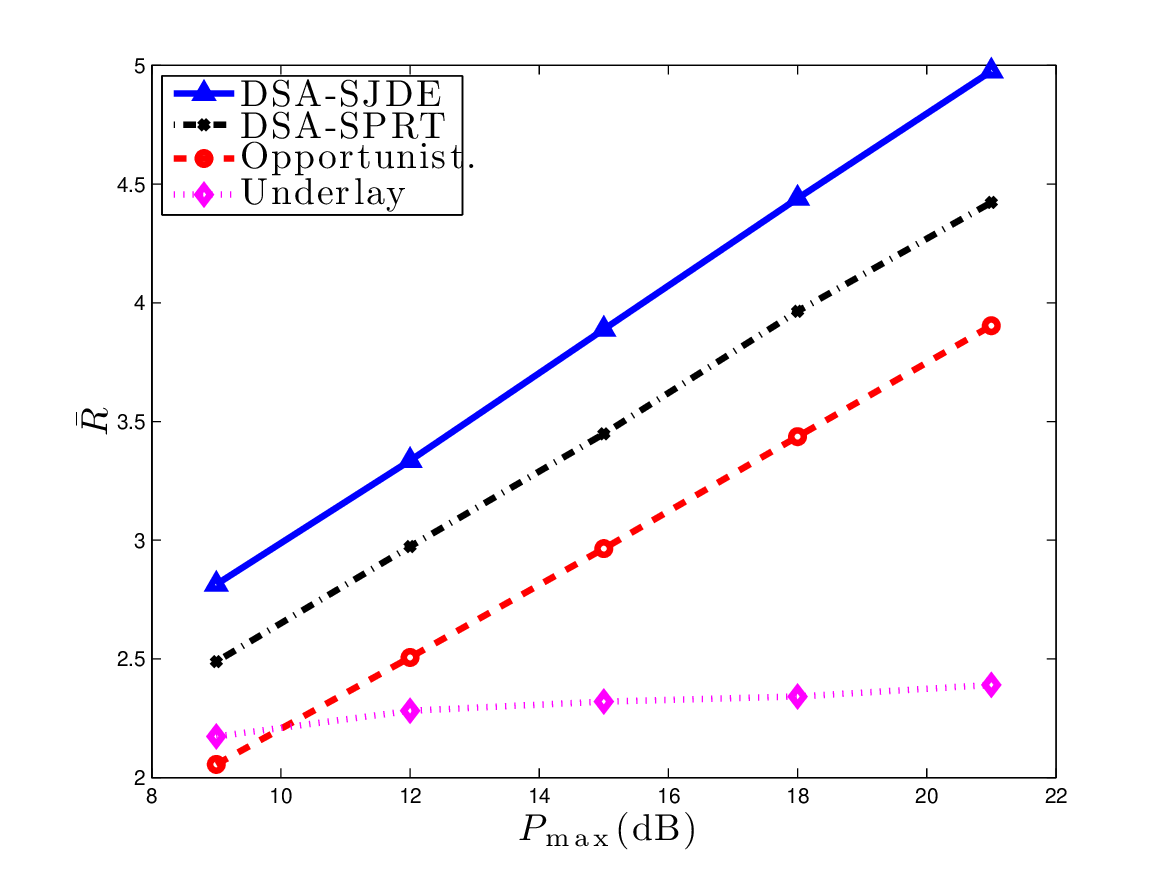}
\caption{Average SU throughput vs. SU maximum power for the conventional (underlay and opportunistic) and the proposed (DSA-SJDE and DSA-SPRT) dynamic spectrum access schemes.}
\label{fig:R_Pmax}
\end{figure}

\underline{{\bf SU throughput vs. SU maximum power}}: We next plot $\bar{R}$ vs. $P_{\text{max}}\in[9~\text{dB},21~\text{dB}]$ in Fig. \ref{fig:R_Pmax}, where $\Pro_{\text{out}}=0.075$, $\pi_0=0.5$, $T=10\times{T_p}$. In this figure, we see that the sensing-based-schemes greatly benefit from increasing $P_{\text{max}}$ as they set their transmit power to $P_{\text{max}}$ when $\Hyp_0$ is decided. In contrast, in the underlay scheme, where no spectrum sensing is performed, the direct effect of increasing $P_{\text{max}}$ is not observed. For small $P_{\text{max}}$ values, the utility of spectrum sensing is deemphasized, and the advantage of the perfect CSI assumption of the underlay scheme becomes apparent. It is again notable that the proposed sequential schemes, especially DSA-SJDE, considerably outperform the conventional methods.

\begin{figure}[t]
\centering
\includegraphics[width=.8\linewidth]{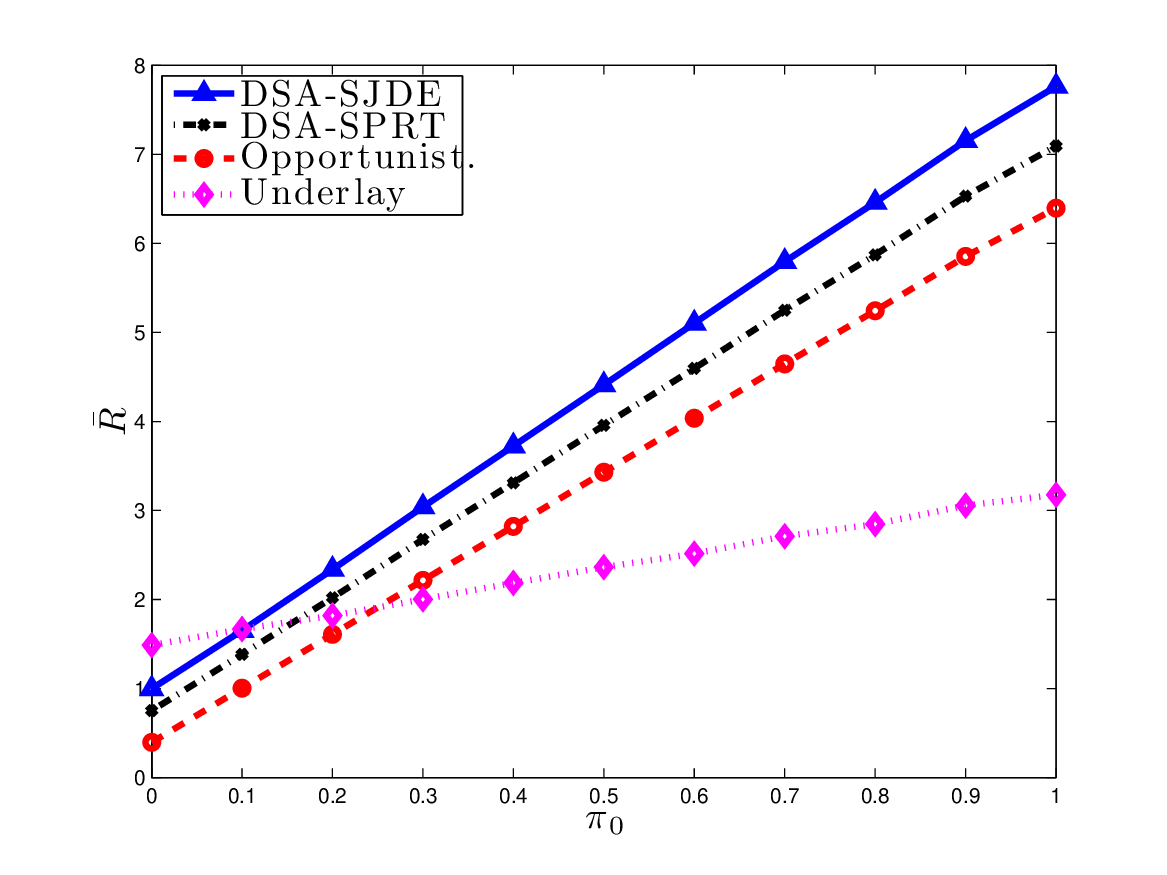}
\caption{Average SU throughput vs. $\Hyp_0$ prior probability for the conventional (underlay and opportunistic) and the proposed (DSA-SJDE and DSA-SPRT) dynamic spectrum access schemes.}
\label{fig:R_P0}
\end{figure}

\underline{{\bf SU throughput vs. $\Hyp_0$ prior probability}}: In the next set of simulations, we investigate the effect of the prior probability $\pi_0$ of $\Hyp_0$ on the average SU throughput, $\bar{R}$, while we set $P_{\text{max}}=15$ dB, $\Pro_{\text{out}}=0.075$, and $T=10\times{T_p}$. Because of the same reason in the changing $P_{\text{max}}$ case the sensing-based-schemes significantly improve their performances with increasing $\pi_0$, as shown in Fig. \ref{fig:R_P0}. The advantage of perfect CSI in the underlay scheme is even more emphasized here, e.g., underlay outperforms the sensing-based-schemes for $\pi_0=0$. The slight improvement in the underlay performance with increasing $\pi_0$ is due to the lack of interference at the SU receiver under $\Hyp_0$.

\begin{figure}[t]
\centering
\includegraphics[width=.8\linewidth]{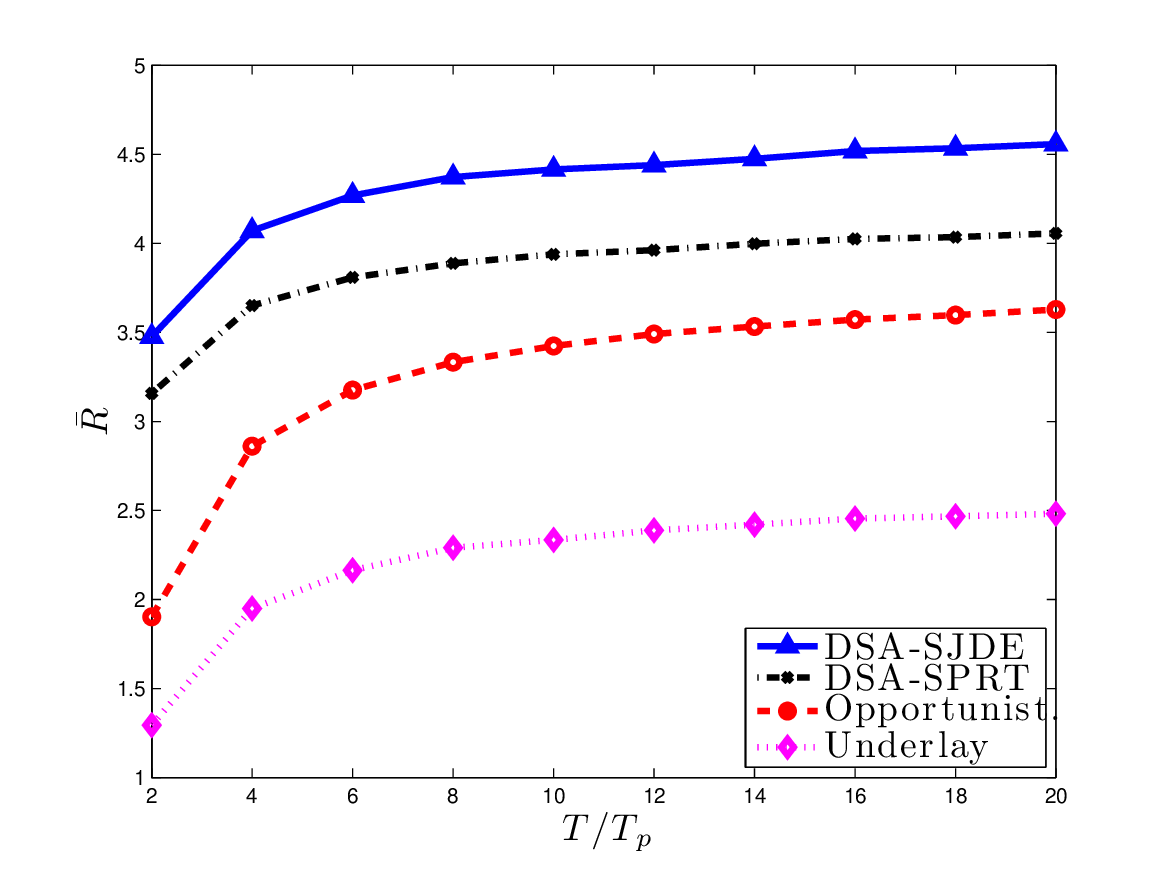}
\caption{Average SU throughput vs. frame length in terms of preamble duration for the conventional (underlay and opportunistic) and the proposed (DSA-SJDE and DSA-SPRT) dynamic spectrum access schemes.}
\label{fig:R_T}
\end{figure}

\underline{{\bf SU throughput vs. Frame length}}: Finally, in Fig. \ref{fig:R_T} we set $P_{\text{max}}=15$ dB, $\Pro_{\text{out}}=0.075$, $\pi_0=0.5$, and analyze the effect of the frame length $T$ on $\bar{R}$. Note that $T$ corresponds to the coherence time in the system. PUs carry out preamble communication every $T$ seconds to estimate the changing channel coefficients. In other words, it is assumed that the channels do not change during each frame of length $T$. Hence, small $T$ corresponds to fast fading channels, whereas large $T$ implies slow fading channels.

Changing $T$ while keeping the preamble duration $T_p$ fixed does not affect the detection and estimation performances, but only changes the remaining time for data transmission, i.e., the scaling term in the $\bar{R}$ expression in \eqref{eq:avg_thr}. Since the scaling term is common to all schemes, they all exhibit similar behaviors with changing $T$. After some certain value, e.g., $T/T_p=10$, the scaling term well approximates unity, and as a result the throughput curves saturate.

\section{Conclusions}
\label{sec:conc}

We have considered dynamic spectrum access under statistical CSI. For a cognitive radio network, a cooperative scheme based on sequential joint spectrum sensing and channel estimation has been proposed. With the objective of SU throughput maximization subject to PU outage constraints, the sensing time needs to be minimized, hence the sequential framework is a better fit to the problem of interest than the fixed-sample-size framework. Unlike the existing works in the literature, channel estimation, which is of practical interest, has been included in the problem formulation. A salient feature of the proposed scheme is that the sensing decision rule makes use of the side information on the true hypothesis provided by the channel estimator. A bandwidth and energy-efficient nonuniform sampling technique, called level-triggered sampling, is used to transmit the information from SUs to the FC, which makes the spectrum sensing decision. Then, the sensing decision and the channel estimates are employed to determine the SU transmit power. Through simulations we have shown the superior performance of the proposed scheme in terms of the average SU throughput over its counterpart that treat the sensing and estimation problems separately, and the conventional spectrum access methods (underlay and opportunistic access) under the same PU outage constraints.

\section*{Appendix: Proof of Theorem \ref{thm:mult}}

As shown in \cite{Yilmaz13_jde} the optimum estimators, decision function, and the stopping time can be found separately, i.e., we can fix two of them, and find the optimum solution for the remaining one.
Furthermore, since $\{h_i^n\}$ are independent, we can minimize each MSE term individually over the corresponding estimator. Hence, the MMSE estimator in \eqref{eq:opt_est_SU} is the optimum estimator for each $h_i^n$. 

Next, substituting the MMSE estimates of $\{h_i^n\}$ into \eqref{eq:cost_mult} we seek the optimum decision rule. From the classical estimation theory (e.g., \cite[page 151]{Poor94}) we know that the conditional mean of the parameter to be estimated gives the MMSE estimator, i.e., $\bExp_1[h_i^n|\cG_t]=\hat{h}_i^n[t]$, and its conditional variance is $\bExp_1\left[\left(h_i^n-\hat{h}_i^n[t]\right)^2\right]=\frac{N_0^i/2}{U_t^i+\frac{N_0^i}{\sigma_i^2}}$. Hence, using
\begin{multline}
  \bExp_1\left[\left(\hat{h}_i^n[\tau]-h_i^n\right)^2 \ind{d_{\tau}=1}\big|\cF_{\tau}\right] = \sum_{t=0}^{\infty} \Exp_1\left[ \frac{N_0^i/2}{U_t^i+\frac{N_0^i}{\sigma_i^2}}\ind{d_t=1}\Big|\cF_t \right] \ind{\tau=t} \\
  = \sum_{t=0}^{\infty} \frac{N_0^i/2}{U_t^i+\frac{N_0^i}{\sigma_i^2}} \Pro_1\left(d_t=1|\cF_t\right) \ind{\tau=t} = \frac{N_0^i/2}{U_{\tau}^i+\frac{N_0^i}{\sigma_i^2}} \Pro_1\left(d_{\tau}=1|\cF_{\tau}\right)
\end{multline}
and
\begin{multline}
  \bExp_1\left[(h_i^n)^2 \ind{d_{\tau}=0} \big|\cF_{\tau}\right] = \sum_{t=0}^{\infty} \bExp_1\left[ (h_i^n)^2 \ind{d_t=0}\big|\cF_t \right] \ind{\tau=t} \\
  = \sum_{t=0}^{\infty} \Exp_1\left[ \bExp_1\left[ (h_i^n)^2|\cG_t \right] \ind{d_t=0}\big|\cF_t \right] \ind{\tau=t} = \Exp_1\left[ \left((\hat{h}_i^n[\tau]\right)^2\ind{d_{\tau}=0}|\cF_t \right] +  \frac{N_0^i/2}{U_{\tau}^i+\frac{N_0^i}{\sigma_i^2}} \Pro_1\left(d_{\tau}=0|\cF_{\tau}\right)
\end{multline}
we can rewrite the cost in \eqref{eq:cost_mult} as
\begin{multline}
\label{eq:cost_mult_dec}
    \cC\left(\tau,d_{\tau}\right) = c_0 \Pro_0\left(d_{\tau}=1|\cF_{\tau}\right) + c_1 \Pro_1\left(d_{\tau}=0|\cF_{\tau}\right) \\ + c_e \sum_{i=1}^2 \sum_{n=1}^2 \left( \Exp_1\left[\left(\hat{h}_i^n[\tau]\right)^2 \ind{d_{\tau}=0}\Big|\cF_{\tau}\right] + \frac{N_0^i/2}{U_{\tau}^i+\frac{N_0^i}{\sigma_i^2}} \right),
\end{multline}
where $\hat{h}_i^n[\tau]$ is given by \eqref{eq:opt_est_SU}. Since the last term in \eqref{eq:cost_mult_dec} does not depend on $d_{\tau}$, we consider only the remaining terms, i.e., 
\be
\label{eq:cost_mult_dec_1}
	\tilde{\cC}\left(\tau,d_{\tau}\right) = c_0 \Pro_0\left(d_{\tau}=1|\cF_{\tau}\right) + c_1 \Pro_1\left(d_{\tau}=0|\cF_{\tau}\right) + c_e \sum_{i=1}^2 \sum_{n=1}^2 \Exp_1\left[\left(\hat{h}_i^n[\tau]\right)^2 \ind{d_{\tau}=0}\Big|\cF_{\tau}\right].
\ee 
We next combine the terms on the right-hand side of \eqref{eq:cost_mult_dec_1} under $\Exp_0$ by changing the measure under $\Hyp_1$ to its counterpart under $\Hyp_0$. The likelihood ratio $\frac{f_1(\left\{y_i^n[t],p_i[t]\right\})}{f_0(\left\{y_i^n[t],p_i[t]\right\})}=e^{L_{\tau}}$ is used for change of measures. 
\begin{align}
\label{eq:cost_mult_dec_0}
    \tilde{\cC}\left(\tau,d_{\tau}\right) =& \Exp_0\left[ c_0 \ind{d_{\tau}=1} + e^{L_{\tau}}\left\{c_1+c_e \sum_{i=1}^2 \sum_{n=1}^2 \left(\hat{h}_i^n[\tau]\right)^2 \right\} \ind{d_{\tau}=0} \Big|\cF_{\tau} \right] \nn\\
    =& \sum_{t=0}^{\infty} \Exp_0\left[ c_0 \ind{d_t=1} + e^{L_t}\left\{c_1+c_e \sum_{i=1}^2 \sum_{n=1}^2 \left(\hat{h}_i^n[t]\right)^2 \right\} \ind{d_t=0} \Big|\cF_t \right] \ind{\tau=t} \nn\\
    =& \sum_{t=0}^{\infty} \Exp_0\left[ \left( c_0 - e^{L_t}\left\{c_1+c_e \sum_{i=1}^2 \sum_{n=1}^2 \left(\hat{h}_i^n[t]\right)^2 \right\}\right) \ind{d_t=1} \Big|\cF_t \right] \ind{\tau=t} \\
    &+ \sum_{t=0}^{\infty} \Exp_0\left[ e^{L_t}\left\{c_1+c_e \sum_{i=1}^2 \sum_{n=1}^2 \left(\hat{h}_i^n[t]\right)^2 \right\} \Big|\cF_t \right] \ind{\tau=t}, \nn
\end{align}
The optimum decision rule that minimizes \eqref{eq:cost_mult_dec_0} selects $\Hyp_1$, i.e., $d_{\tau}=1$, when $$c_0 \leq e^{L_t}\left\{c_1+c_e \sum_{i=1}^2 \sum_{n=1}^2 \left(\hat{h}_i^n[\tau]\right)^2 \right\},$$ and selects $\Hyp_0$ otherwise, proving \eqref{eq:opt_dec_SU}. 

Finally, substituting the optimum detector into the cost function \eqref{eq:cost_mult_dec} we have
\begin{multline}
\label{eq:cost_mult_dec_st}
	\cC\left(\tau\right) = \Exp_0\left[ \left( c_0 - e^{L_{\tau}}\left\{c_1+c_e \sum_{i=1}^2 \sum_{n=1}^2 \left(\hat{h}_i^n[\tau]\right)^2 \right\}\right)^- \Big|\cF_{\tau} \right] + c_1 \\
	+ c_e \sum_{i=1}^2 \sum_{n=1}^2 \left( \Exp_1\left[ \left(\hat{h}_i^n[\tau]\right)^2 \big|\cF_{\tau} \right] + \frac{N_0^i/2}{U_{\tau}^i+\frac{N_0^i}{\sigma_i^2}} \right),
\end{multline}
where $(x)^-=\min(x,0)$ is the negative part operator.
We now focus on $\Exp_1\left[ \left(\hat{h}_i^n[t]\right)^2 \big|\cF_{\tau} \right]$, where $\hat{h}_i^n[t] = \frac{V_t^{in}+\frac{\mu_i}{2}\frac{N_0^i}{\sigma_i^2}} {U_t^i+\frac{N_0^i}{\sigma_i^2}}$ is given by \eqref{eq:opt_est_SU}. Note from \eqref{eq:signal_model_real} that under $\Hyp_1$ given $\cF_t$ we have $V_t^{in}=\sum_{m=1}^t y_i^n[m] \sim \cN\left(\frac{\mu_i}{2}U_t^i,\frac{\sigma_i^2}{2}\sum_{m=1}^t|p_i[m]|^4 + \frac{N_0^i}{2}U_t^i\right)$, hence $\hat{h}_i^n[t]$ is Gaussian with mean $\frac{\mu_i}{2}$ and variance $\frac{\frac{\sigma_i^2}{2}\sum_{m=1}^t|p_i[m]|^4 + \frac{N_0^i}{2}U_t^i}{\left(U_t^i+\frac{N_0^i}{\sigma_i^2}\right)^2}$. Therefore, $\Exp_1\left[ \left(\hat{h}_i^n[t]\right)^2 \big|\cF_{\tau} \right] = \frac{\mu_i^2}{4}+\frac{\frac{\sigma_i^2}{2}\sum_{m=1}^t|p_i[m]|^4 + \frac{N_0^i}{2}U_t^i}{\left(U_t^i+\frac{N_0^i}{\sigma_i^2}\right)^2}$, which is decreasing in $U_t^i$. As a result, the last term in \eqref{eq:cost_mult_dec_st} is decreasing in $U_t$. Indeed the first term is also decreasing in $U_t$, hence the optimum stopping rule is a thresholding on the conditional Fisher information $U_t$ as shown in \eqref{eq:opt_stop_SU}. The analysis of the first term, which is very technical and involved, directly follows from \cite[Theorem 1]{Yilmaz13_jde}, thus is omitted here.

\end{document}